# Dielectrophoretic response and electro-deformation of soft bioparticles interacting with a metallo-dielectric Janus active particle

*Donggang Cao[1], and Gilad Yossifon[1,2]\**

[1]School of Mechanical Engineering, Tel-Aviv University, Tel-Aviv, 69978, Israel

[2]Department of Biomedical Engineering, Tel-Aviv University, Tel-Aviv, 69978, Israel

\*E-mail: gyossifon@tauex.tau.ac.il

Active (self-propelling) particles have emerged as innovative microscale tools in the field of single cell analysis with the advantages of being untethered, remotely controlled, hybrid powered, with sub-cellular precision. This study investigates the dielectrophoretic (DEP) response and electro-mechanical deformation of cell nuclei interacting with active metallo-dielectric Janus Particles (JPs) under an externally applied electric field. An "equivalent droplet" two-phase model is employed to simulate the bioparticle, coupling the Navier-Stokes equations with the Phase Field Model to capture fluid motion and interface dynamics. Good qualitative agreement is obtained among experimental, analytical, and numerical results. The findings reveal a nonlinear relationship between nucleus deformation and its surface coverage of the JP with respect to the applied voltage. The overall coverage ratio of the JP's dielectric hemisphere increases with voltage as the positive DEP force on the dielectric side strengthens, exhibiting maximum at a certain voltage. The strong correlation between nucleus flexibility and JP surface coverage suggests that the JP coverage ratio could serve as a biomechanical marker for nucleus deformability, providing a novel method for in-situ evaluation of nucleus mechanics.

## 1. Introduction

Self-propelling micromotors have garnered significant attention in recent years for their potential as innovative microscale tools in single-cell analysis [1–3]. These micromotors enable precise manipulation [4] and probing of individual cells [5] and subcellular components [6], advancing various biomedical applications [7,8]. Among the diverse range of micromotors and powering mechanisms [9-14], electrically powered metallo-dielectric Janus particles (JPs) have emerged as particularly useful due to their simple fabrication and the advantages of electrical actuation—being fuel-free, label-free, biocompatible, and easily controlled by adjusting the electric field magnitude and frequency. These properties enable complex electrokinetic JP actuations [15-17] and make them particularly advantageous for label-free dielectrophoretic (DEP) manipulations, where DEP forces [18,19] and electro-deformation effects [20,21] can be harnessed to manipulate soft bioparticles, such as cell nuclei [22], with high precision. These capabilities open new possibilities in targeted drug delivery [23], cell sorting [24], and single-cell mechanical characterization [25].

The asymmetric design of JPs allows them to function as mobile microelectrodes under an applied electric field, generating localized field gradients that drive DEP forces [26]. These forces can trap and deform soft bioparticles, providing a dual mechanism for both manipulation and mechanical assessment [27]. The electro-deformation of trapped cells and nuclei under DEP forces offers valuable insights into cellular mechanics, which are critical for understanding cellular health, disease mechanisms, and mechanical phenotyping—the assessment of structural integrity and mechanical properties in both healthy and diseased cells [28,29].

While experimental studies have provided valuable preliminary insights into JP-bioparticle interactions, intriguing phenomena such as JP insertion and engulfment at cell membranes have been observed [30,31]. Our recent work has demonstrated that active, polarizable metallo-dielectric JPs effectively function as mobile microelectrodes, generating the necessary field gradients for efficient DEP manipulation [32]. Furthermore, we have highlighted the strong correlation between DEP-induced electro-deformation of cell nuclei and applied field strength, underscoring the potential of JPs as active carriers for mechanical probing of subcellular components [33]. However, to the best of our knowledge, no systematic simulation studies have been conducted to investigate these complex JP-bioparticle interactions, particularly in DEP-based applications. This gap hinders the comprehensive interpretation of experimental data and the optimization of JP-bioparticle interactions for practical biomedical use. To address this gap, the present study employs numerical simulations using COMSOL Multiphysics 6.2 [34] to analyse the DEP-driven interactions between JPs and deformable particles, with a particular focus on cell nuclei. An "equivalent droplet" two-phase model is employed to simulate the deformation of soft bioparticles under varying electric fields [35-36] (see Section 4 for details). This model integrates the Navier-Stokes equations with a multiphase approach to accurately capture fluid dynamics and interface behaviour. The present study systematically investigates how particle composition, applied voltage, and nucleus stiffness influence dielectrophoretic response and electro-deformation. By providing a comprehensive framework for understanding these interactions, this research enhances fundamental physical insights, predictive modelling, and the design of DEP-based biomanipulation strategies, with broad implications for active particle-based biomedical applications.

## 2. Results and Discussion

### 2.1. Overview of JP-nucleus interaction

Our previously studied experimental setup [33] consists of a simple microfluidic chamber, formed by positioning a 120μm-high space between two parallel indium tin oxide (ITO)-coated glass substrates. JPs consisting of gold half-coated polystyrene spherical particles, are introduced into the chamber along with target cell nuclei. The manipulation and transport of the JPs are controlled by an externally applied electric field, as schematically illustrated in Figure 1A. Figure 1B schematically represents the nucleus as a homogeneous "equivalent droplet," incorporating the combined properties of the nucleus's cytoplasm and membrane [37-39]. The simulations utilize the Phase Field Model [40-42] coupled with the Navier-Stokes equations to capture the interface dynamics between the continuous and dispersed phases. The electric force is computed using the divergence of the Maxwell stress tensor, while the electric field is resolved through the current conservation equation. These simulations account for the interplay between hydrodynamic and electric forces acting on the nucleus, leading to its movement and deformation.

When exposed to an electric field, the JP attracts and deforms the nucleus onto its surface, enabling it to carry and transport the loaded nucleus [33] (Figure 1C). Figure 1D shows numerical simulation snapshots illustrating the JP-nucleus interaction over time. The nucleus is attracted to the JP by DEP forces, gradually deforming and migrating until it reaches a steady-state shape, settling on the JP's surface and covering parts of both the gold and polystyrene hemispheres. This behaviour aligns qualitatively with experimental observations (see Movie 1 in the Supporting Information).

Figure 1E presents the time-evolving normalized electric field distribution, further supporting the experimental findings. Both nucleus and JP undergo polarization in the electric field, with the JP inducing perturbations that generate distinct dielectrophoretic (DEP) potential wells, labelled ①-⑥ in the first panel of Figure 1E. Local maxima in the electric field at locations ① and ⑤ indicate regions where positive DEP (pDEP) trapping of bioparticles is likely, while a weaker pDEP well appears on the polystyrene side at location ④. Conversely, locations ②, ③, and ⑥ correspond to negative DEP (nDEP) trapping sites due to local electric field minima. As a homogeneous dielectric droplet exhibiting a pDEP response [33], the nucleus migrates toward the high-intensity potential well at location ①. Initially, the nucleus moves toward the conductive hemisphere at location ① due to pDEP attraction. Upon contact with the JP, the electric field intensity at the JP-nucleus interface (location ①) gradually decreases until the system reaches a steady-state deformed nucleus configuration.

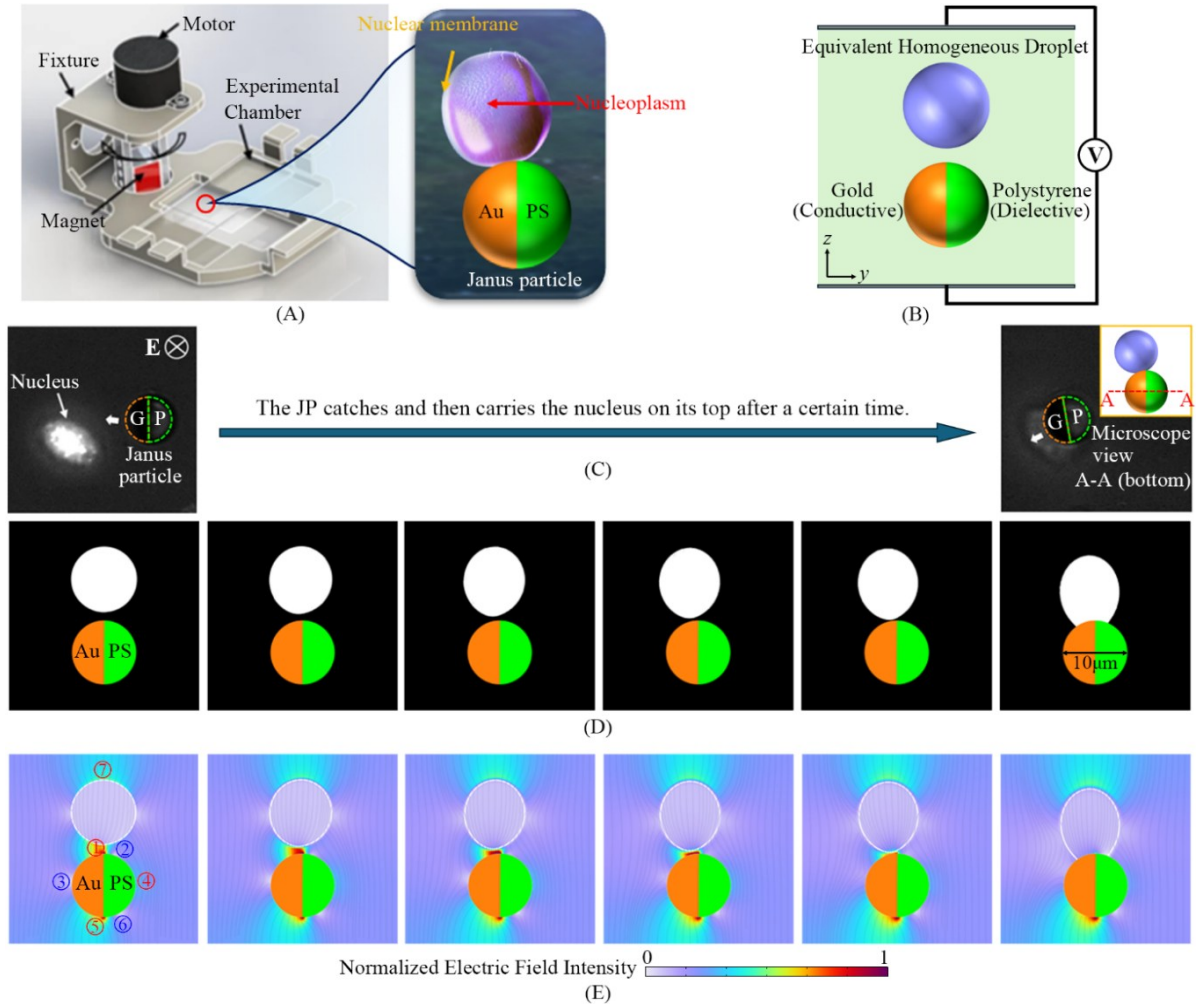

**Figure 1.** Experimental and numerical system setups and overview of JP-nucleus interactions: (A) Schematic of the experimental setup [33]. (B) Schematic representation of the simulation model. (C) Time evolution of JP-nucleus interaction observed experimentally, showing nucleus attraction and deformation onto the JP (inset: side-view schematic) [33]. (D) Corresponding numerical simulation snapshots. (E) Normalized electric field intensity distributions at different time points (white line indicates the nucleus surface contour).

## 2.2. Interactions between nuclei and different particles

The spatial variation of induced dielectrophoretic (DEP) potential wells on the JP's surface plays a crucial role in governing JP-nucleus interactions, depending on their relative positioning. To further investigate this, we conducted a comparative numerical analysis of interactions involving four nuclei positioned around three types of particles: homogeneously conductive, homogeneously dielectric, and Janus (Figure 2). The conductive gold particle exhibits positive

DEP trapping at the top and bottom, while negative DEP regions appear on its lateral sides, whereas the dielectric particle shows the opposite behaviour. Consequently, nuclei above or below the gold particle are attracted to it, while those on the lateral sides experience repulsion, and vice versa for the dielectric particle. Interactions with the JP combines the effects of both the gold (conductive) and dielectric particles. Nuclei located above or below the JP predominantly migrate toward the conductive side due to positive DEP, while nuclei adjacent to the conductive lateral surface are repelled, and those near the dielectric side are attracted. The hybrid nature of the JP thus enables complex, spatially selective interactions with target nuclei.

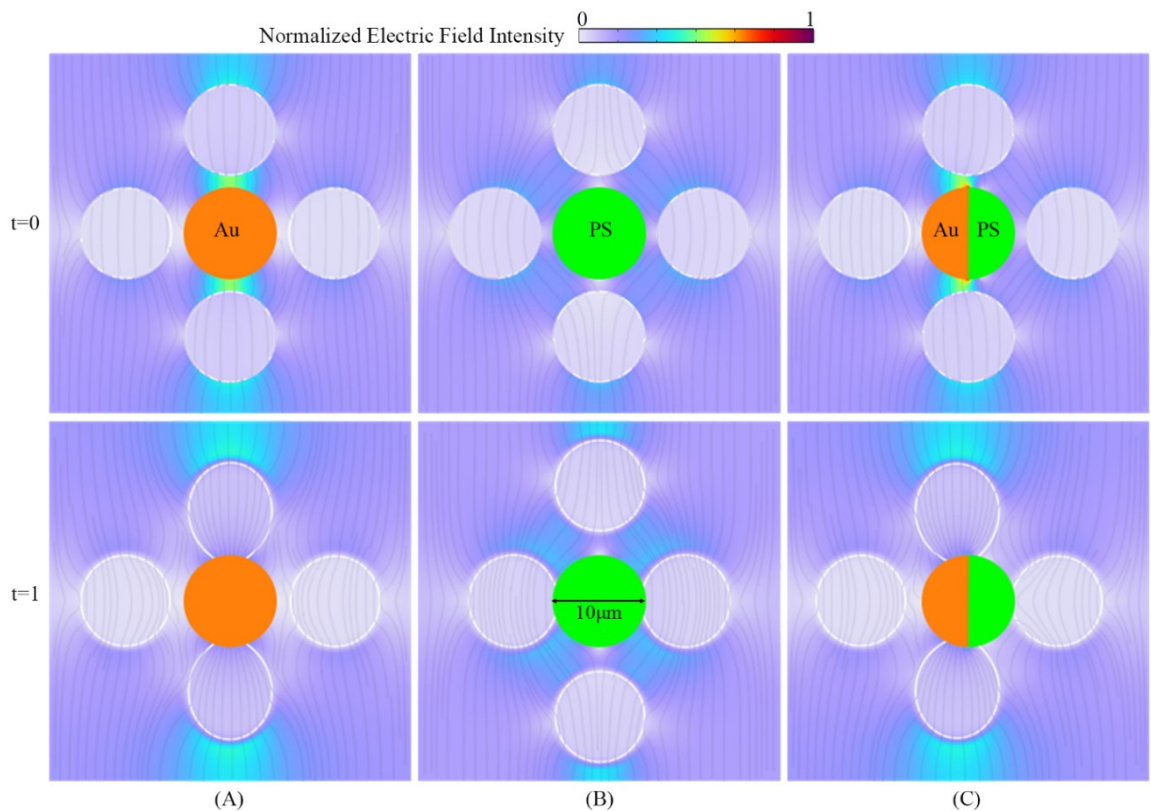

**Figure 2.** Normalized electric field distributions (scaled by $10^6$V/m) are shown for nuclei positioned at various locations around (A) a conductive particle, (B) a dielectric particle, and (C) a Janus particle at different normalized time points (scaled by 15ms).

## 2.3. Impact of the applied electric field on the Janus-nucleus interaction

Our previous experimental study [33] demonstrated that a nucleus trapped by a JP can spread onto the polystyrene (PS) hemisphere, with the extent of coverage directly correlated to the applied voltage. Figure 3 provides a comprehensive analysis of this interaction, comparing experimental observations [33] with numerical simulations. Both 2D and 3D models are employed to simulate the JP-nucleus interaction. As shown in Figure 3A, results from the central plane of interest (plane A-A) in the 3D simulations closely align with those from the 2D simulations, confirming the viability of the 2D approach. Therefore, given the high computational cost and time demands of 3D simulations, we propose using the more efficient qualitative 2D simulations to systematically investigate the impact of the applied electric field on JP-nucleus interactions. Figure 3B shows experimental images [33] of the nucleus (in red) and the JP surface (outlined by a dotted yellow circle) at 4V and 18V. The increasing coverage of the JP's PS hemisphere by the nucleus (shaded sector) indicates enhanced electro-deformation under higher electric fields. The corresponding simulation snapshots in Figure 3C further support this trend, showing increased nucleus coverage on the PS hemisphere as voltage rises. The strong qualitative agreement between experimental and numerical results validates the simulation approach, reinforcing its utility for further analysis.

Figure 3D illustrates the electric field magnitude distribution around the JP and nucleus at different voltages, demonstrating how the field evolves as the applied voltage increases. As described in Section 2.1, the nucleus experiences pDEP forces, migrating toward regions of higher electric field intensity. Initially, the strongest field intensity at location ① serves as the primary trapping site. However, as the nucleus is captured, this pDEP well weakens, and other wells, particularly on the PS side (location ④), become more influential. The increasing field intensity at ④ at higher voltages enhances the nucleus's attraction to the polystyrene hemisphere.

The quantitative analysis of the dielectrophoretic coverage ratio (Figure 3E), defined as the fraction of the JP's PS hemisphere covered by the nucleus ($\eta_1 = (\theta_1/\pi) \times 100\%$, where $\theta_1$ is the angular coverage), further supports the validity of the 2D simulations. Despite the limited number of 3D data points due to computational constraints, both 2D and 3D simulations align well with experimental results. Generally, the coverage ratio increases with the applied voltage up to approximately 14V, after which a slight decline is observed. This non-monotonic behaviour may result from competing mechanisms: on one hand, increased electric field intensity enhances pDEP-driven attraction toward location ④, promoting greater nucleus coverage of the dielectric side; on the other hand, elongation of the nucleus along the field lines reduces its effective contact area with the JP, thereby decreasing $\theta_1$. This elongation also influences the total coverage ratio ($\eta_2 = (\theta_2/2\pi) \times 100\%$), which declines with increased voltages (Figure 3F). Although the 2D simulations qualitatively match the 3D results on the plane of interest in terms of coverage ratios, they do not fully replicate the deformation ratio of the 3D system (Figure 3G). The deformation ratio, defined as the percentage increase in the nucleus's surface area ($\varphi = (\Delta S/S) \times 100\%$), quantifies how much the nucleus stretches under the electric field. Since 2D simulations inherently capture deformation in only two directions, the deformation ratios are consistently lower than those from 3D simulations. Hence, while the 2D simulations have quantitative limitations in predicting deformation ratios, they provide valuable insights into the key mechanisms of JP-nucleus interaction.

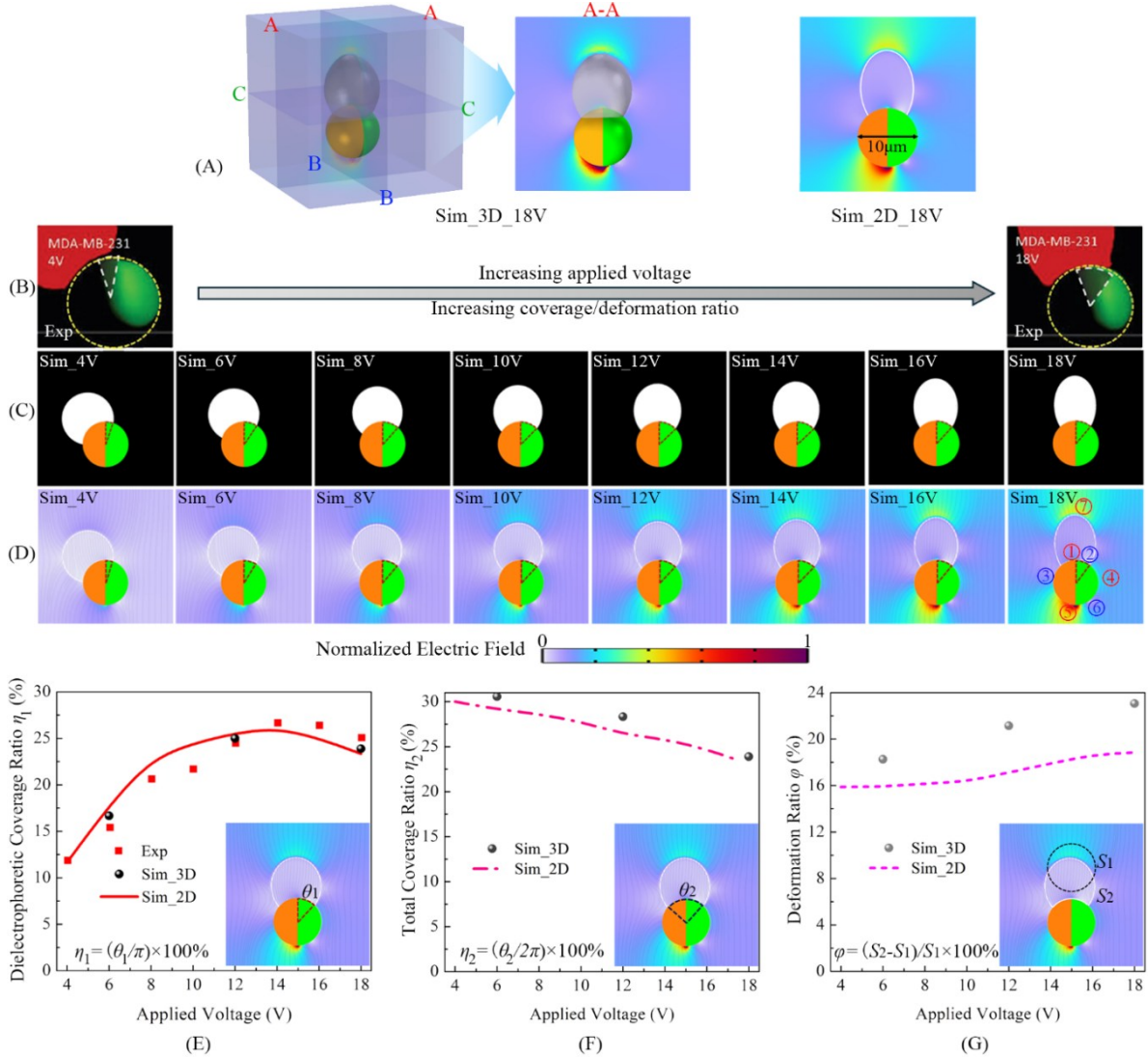

**Figure 3.** JP-Nucleus interaction under varying electric fields at 1MHz: (A) Comparison of 3D and 2D simulation results; (B) Experimental images of the nucleus (red) and Janus particle (JP, dotted yellow outline) reconstructed using Imaris 5.0 at the central plane. The green region represents the polystyrene (PS) hemisphere of the JP, while the shaded circular sector denotes the PS portion covered by the nucleus [33]; (C) Numerical images of the nucleus (white) and JP, with the dashed circular sector indicating the PS hemisphere region covered by the nucleus; (D) Simulated electric field distributions during JP-nucleus interactions at different applied voltages; (E) Experimental and simulated dielectrophoretic coverage ratio of the PS hemisphere versus applied voltage; (F) Total coverage ratio (including both PS and Au hemispheres) versus applied voltage; (G) Deformation ratio of the nucleus versus applied voltage.

**2.4. Impact of the nuclei stiffness on the Janus-nucleus interaction**

Nucleus deformability, or stiffness, plays a critical role in determining the interaction dynamics between a nucleus and a JP under an applied electric field. To investigate this effect, we examined the interactions of a JP with two cell nuclei of differing stiffnesses: the highly deformable MDA-MB-231 nucleus and the more rigid MCF-7 nucleus. This distinction is biologically significant, as MDA-MB-231 cells exhibit a higher metastatic potential than MCF-7 cells [43]. To quantify nucleus deformability, the transmigration-based deformability assay (TDA) using Transwell inserts with 5 μm and 8 μm pores was employed in Ref. [33]. In this assay, MDA-MB-231 cells - known for their high deformability - exhibited increased transmigration, with over 180 cells passing through both pore sizes. This high rate of passage suggests that their nuclei can easily deform and squeeze through narrow openings. In contrast, MCF-7 cells, with lower deformability, depicts limited transmigration, with only about 50 cells passing through the 8 μm pores and very few through the 5 μm pores, highlighting the stiffness of their nuclei (Figure 4A).

Correspondingly, simulations have been conducted by modelling the nucleus deformability through an effective surface tension parameter ($\gamma$). Figure 4B presents experimental images and simulation snapshots comparing interactions of the JP with nuclei of varying deformability (represented by the normalized equivalent surface tension, $\Gamma = \gamma/\gamma_o$, with $\gamma_o = 0.0002$ N/m [44]) at 4V and 18V. For the highly deformable MDA-MB-231 nuclei, simulations with lower $\Gamma$ values (indicating higher deformability) show increased coverage of the JP's polystyrene side. Conversely, stiffer MCF-7 nuclei exhibit reduced coverage, in qualitative agreement with experimental observations. Figure 4C quantifies the dielectrophoretic coverage ratio ($\eta_1$) as a function of applied voltage for nuclei with varying surface tension values. Across all deformability levels, $\eta_1$ increases with voltage, but the rate of increase is strongly influenced by nucleus stiffness. Softer nuclei (lower $\Gamma$ values, e.g., MDA-MB-231) exhibit a steep rise in $\eta_1$, indicating their greater ability to conform to the JP's surface under stronger electric fields. Conversely, stiffer nuclei (higher $\Gamma$ values, e.g., MCF-7) show a more gradual and limited increase in $\eta_1$, reflecting their resistance to deformation. All curves follow a similar trend: an initial rapid rise in $\eta_1$, followed by a reduced growth rate beyond a critical transition voltage. This transition voltage shifts higher for stiffer nuclei, highlighting the interplay between deformability and electric field strength.

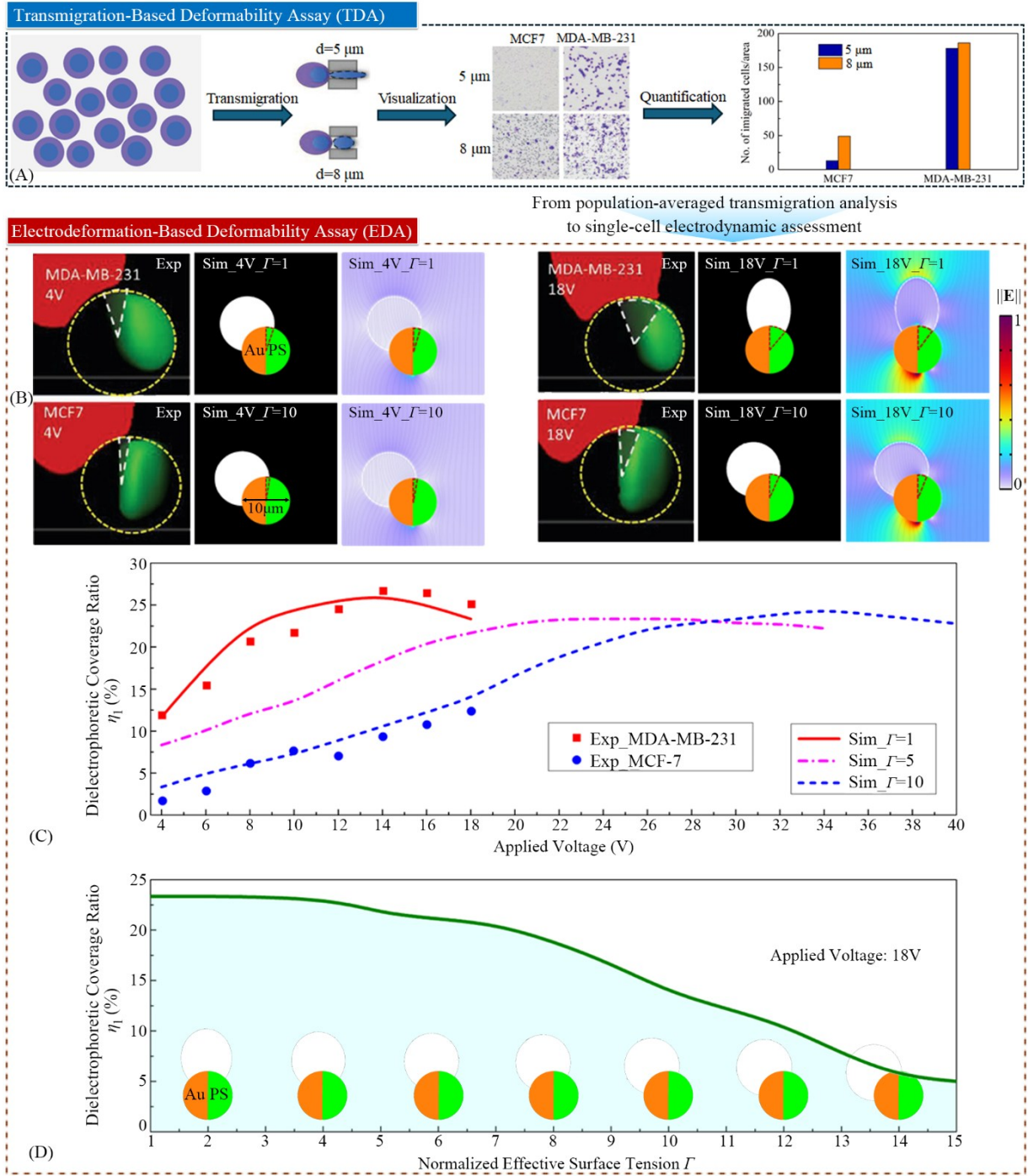

**Figure 4.** Influence of nucleus stiffness on JP-nucleus interaction at 1MHz: (A) Schematic of the transmigration-based deformability assay (TDA) and results from Ref. [33] demonstrating the deformability of MDA-MB-231 and MCF-7 cells; (B) Comparison of experimental [33] and numerical results for JP-nucleus interactions with varying nucleus stiffness (MDA-MB-231 and MCF-7) at 4V and 18V; (C) Simulated dielectrophoretic coverage ratio as a function of applied voltage for different nucleus stiffnesses (represented by equivalent surface tension, $\Gamma$); (D) Simulated dielectrophoretic coverage ratio as a function of equivalent surface tension ($\Gamma$) at the applied voltage of 18V.

Figure 4D further explores the relationship between $\eta_1$ and $\Gamma$ at 18V and 1MHz. The inverse correlation confirms that stiffer nuclei (higher $\Gamma$) achieve significantly lower dielectrophoretic coverage ratios due to their rigidity. These results suggest that the JP coverage ratio could serve as a biomechanical marker of nucleus deformability, offering a novel approach for assessing nucleus mechanics. Unlike the population-based TDA assay, which relies on mechanical squeezing forces, our method leverages electrostatic forces and enables single cell resolution.

## 3. Conclusion

This study investigated the DEP response and electro-deformation of soft bioparticles, specifically cell nuclei, interacting with a metallo-dielectric JP under varying electric fields. Using numerical simulations validated against previous experimental observations, we analysed how applied voltage and nucleus deformability influence DEP-induced deformation and the extent of nucleus coverage on the JP surface. Our findings reveal a nonlinear relationship between the dielectrophoretic coverage ratio of the deformable nucleus on the JP's polystyrene hemisphere and the applied voltage. Initially, the coverage ratio increases with voltage, reaching a maximum before slightly declining at higher voltages. This behaviour arises from the dynamic interplay between the soft nucleus and the multiple potential wells formed around the JP's surface. At lower voltages, the strongest positive DEP (pDEP) well at the JP interface attracts and traps the nucleus. However, once the conductive nucleus attaches, this potential well weakens, allowing other pDEP wells—particularly the one on the polystyrene side—to dominate, pulling the nucleus in that direction and stretching it. Nucleus deformability plays a crucial role in modulating the JP-nucleus interaction. Highly deformable nuclei (e.g., MDA-MB-231) exhibit greater coverage and adapt more readily to the JP surface, whereas stiffer nuclei (e.g., MCF-7) show limited interaction due to their reduced ability to deform. These findings highlight the potential of JPs as microscale tools for probing nucleus deformability—a key mechanical biomarker associated with different cellular states and diseases. Since nucleus stiffness strongly influences electro-deformation, it enables selective manipulation and sorting of cells and their nuclei by controlling their loading and positioning on the JP surface. These findings offer valuable guidance for utilizing DEP forces in biomedical applications such as single-cell and organelle mechanical probing, sensing, and selective manipulation based on mechanical and dielectric properties.

## 4. Numerical simulation methodology

To simulate the JP-Nucleus interaction, the incompressible Navier-Stokes equations are coupled with a multiphase framework to resolve fluid dynamics and track interface evolution under electric fields. In the analyzed microfluidic configuration, both continuous and dispersed phases exhibit laminar flow, with negligible influence from buoyancy and gravitational forces. The governing equations, incorporating electric stresses and surface tension, are formulated as:

$$\begin{cases} \rho \nabla \cdot \mathbf{u} = 0 \\ \rho \frac{\partial \mathbf{u}}{\partial t} = \nabla \cdot [-p\mathbf{I} + \mu(\nabla \mathbf{u} + (\nabla \mathbf{u})^T)] + \mathbf{F}_{st} + \mathbf{F}_e \end{cases} \quad (1)$$

where $\mathbf{u}$, $\rho$, $\mu$, $p$ denote the velocity vector, density, dynamic viscosity, and pressure, respectively. $\mathbf{F}_{st}$ represents the surface tension force, and $\mathbf{F}_e$ donates the Maxwell electric force.

To track the interface during the transient evolution of two-phase laminar flow, the Phase-Field method is employed for its efficiency in accurately capturing interface deformation under complex force interactions [45-47]:

$$\begin{cases} \frac{\partial \phi}{\partial t} + \mathbf{u} \cdot \nabla \phi = \nabla \cdot \frac{3\chi\gamma\xi}{2\sqrt{2}} \nabla \psi \\ \psi = -\nabla \cdot \varepsilon^2 \nabla \phi + (\phi^2 - 1)\phi \end{cases} \quad (2)$$

where $\phi$ is the Phase-Field variable, and it is -1 in the dispersed phase and 1 in the continuous phase, $\gamma$ denotes the surface tension coefficient. The numerical parameter, $\xi$, determines the thickness of the fluid interface where the Phase-Field variable $\phi$ varies from -1 to 1 smoothly, and $\chi$ controls the mobility of the interface.

The density and viscosity are automatically calculated from the Phase-Field variable $\phi$:

$$\begin{cases} \rho = \rho_n f_n + \rho_s f_s \\ \mu = \mu_n f_n + \mu_s f_s \end{cases} \quad (3)$$

where the subscripts "*n*" and "*s*" indicate the nucleus and the solution, respectively. The volume fractions of the nucleus phase are $f_n = (1 - \phi)/2$, and the volume fraction of the solution phase is $f_s = (1 + \phi)/2$, which satisfies $f_n + f_s = 1$. The surface tension force is calculated as:

$$\begin{cases} \mathbf{F}_{st} = G\nabla\phi \\ G = \frac{3\gamma\xi}{2\sqrt{2}}\left[-\nabla^2\phi + \frac{(\phi^2-1)\phi}{\xi^2}\right] \end{cases} \quad (4)$$

The electric force is given by the divergence of the Maxwell stress tensor by the equation below [48-49]:

$$\begin{cases} \mathbf{F}_e = \nabla \cdot \mathbf{T} \\ \mathbf{T} = \mathbf{E}\mathbf{D}^T - [(\mathbf{E} \cdot \mathbf{D})\mathbf{I}]/2 \end{cases} \quad (5)$$

wherein $\mathbf{E}$ is the electric field and $\mathbf{D}$ is the electric displacement field with the following expressions:

$$\begin{cases} \mathbf{E} = -\nabla V \\ \mathbf{D} = \varepsilon \mathbf{E} \end{cases}. \tag{6}$$

Furthermore, the current conservation equation can be solved to determine **E**:

$$\begin{cases} \nabla \cdot \left(\varepsilon \frac{\partial \mathbf{E}}{\partial t} + \sigma \mathbf{E}\right) = 0 \\ \varepsilon = \varepsilon_n f_n + \varepsilon_s f_s \\ \sigma = \sigma_n f_n + \sigma_s f_s \end{cases}, \tag{7}$$

where $\varepsilon$ is the permittivity and $\sigma$ is the conductivity.

Directly solving these equations in the physical time domain to simulate nucleus deformation at 1 MHz requires prohibitively small time steps and incurs high computational costs. To balance efficiency and accuracy, the electrical problem is solved in the frequency domain, and a cycle-averaged electric force is applied in the time domain [50]. This is an appropriate approximation given the significant disparity in timescales between fluid motion (~ms) and electrical excitation (~MHz). Consequently, the governing equation for the electric field in the frequency domain becomes:

$$\begin{cases} \nabla \cdot \left(i\omega\varepsilon \tilde{\mathbf{E}} + \sigma \tilde{\mathbf{E}}\right) = 0 \\ \varepsilon = \varepsilon_n f_n + \varepsilon_s f_s \\ \sigma = \sigma_n f_n + \sigma_s f_s \end{cases}, \tag{8}$$

while the cycle-averaged force is formulated as [51]:

$$\begin{cases} \langle \mathbf{F}_e \rangle = \nabla \cdot \langle \mathbf{T} \rangle \\ \langle \mathbf{T} \rangle = 0.25 \mathrm{Re}(\tilde{\varepsilon}) \left[ \left(\tilde{\mathbf{E}}\tilde{\mathbf{E}}^* + \tilde{\mathbf{E}}^*\tilde{\mathbf{E}}\right) - \|\tilde{\mathbf{E}}\|^2 \mathbf{I} \right] \end{cases}. \tag{9}$$

Here, the tilde "~" denotes complex phasor quantities. Only the real part of the medium complex permittivity $\mathrm{Re}(\tilde{\varepsilon}) = \varepsilon$ appears in the stress tensor.

The cell nucleus is modelled as a homogenized "equivalent droplet" with effective electrical properties that capture the combined influence of the nucleus membrane and cytoplasm, as illustrated in Figure S1. The characteristic dielectric properties of the nucleus membrane and cytoplasm, obtained from the literature, are summarized in Table S1. The DEP response across a broad frequency range, evaluated using the Clausius-Mossotti factor, is presented in Figure S2. Table S2 provides the physical properties of the nucleus and the surrounding fluid, selected to satisfy the experimental condition that a positive DEP (pDEP) force arises at the experimentally investigated frequency of 1MHz.

The governing equations are solved with the following boundary conditions: the upper and lower walls act as electrodes with applied voltages, while the left and right channel edges are set to electric insulation. Since our primary focus is on simulating the high frequency domain (~1MHz), consistent with the preliminary experiments [33], we neglect induced-charge electro-osmosis (ICEO) effects and the effective Helmholtz-Smhoulokowski slip boundary condition on the metal hemisphere. This simplification is justified as the operating frequency is significantly higher than the RC frequency associated with the formation of the induced electrical double layer (EDL) ($f_{RC}$=1/2$\pi\tau$=1.3kHz, where $\tau$=$\lambda a/D$ is the induced charge relaxation time; $a$=5μm is the radius of JP; $\lambda = \sqrt{\varepsilon D/\sigma}$ is the Debye length; $\sigma$=0.6mS/m is the solution conductivity; $\varepsilon = 78\varepsilon_0$ is the solution permittivity and $\varepsilon_0$=8.854187817×10$^{-12}$F/m is the vacuum permittivity; $D$=2×10$^{-9}$m$^2$/s is the diffusion coefficient of the ionic species [52]). Accordingly, the JP is modeled by assigning distinct boundary conditions to its two hemispheres: a floating electrode condition for the metallic (gold) hemisphere and an insulating condition for the dielectric (polystyrene) hemisphere [32]. Validation results in the Supporting Information confirm that the applied boundary conditions effectively capture JP–nucleus interactions at 1 MHz (see Figures S3–S4). Additionally, Figure S5 presents the Clausius–Mossotti factor and the corresponding JP–nucleus interaction, derived from the parameters in Table S2 and simulations using the floating boundary condition. These results indicate that, as frequency increases beyond 1 MHz—where positive dielectrophoresis (pDEP) is observed experimentally—a transition to negative DEP (nDEP) occurs, leading the nucleus to migrate toward regions of lower electric field strength. The strong agreement between analytical predictions and numerical simulations further supports the validity and robustness of the proposed modeling approach.

In addition, no-slip boundary conditions are imposed on all solid walls, which is justified at the examined frequency (~1MHz), well above the RC frequency $f_{RC}$=1/2$\pi\tau$=1.3kHz. To accurately resolve the complex interactions in the system, second-order numerical schemes are employed in conjunction with a locally refined mesh (see Figure S6), concentrating computational effort in regions with steep field gradients. Four computational grids with varying cell counts (see Table S3) are used to assess numerical errors arising from limited spatial resolution. The results confirm that a fine mesh with approximately 30,000 cells provides accurate and grid-independent solutions. Further details on the grid-independence validation are provided in Figure S7 of the Supporting Information.

**Supporting Information**

Supporting Information is available from the Wiley Online Library or from the author.

**Acknowledgments**

G.Y. acknowledges support from the Israel Science Foundation (ISF) (1934/20). D.C. is grateful to Anna Juhasz from COMSOL support for her technical support.

Supporting Information

# Dielectrophoretic response and electro-deformation of soft bioparticles interacting with a metallo-dielectric Janus active particle

*Donggang Cao[1], and Gilad Yossifon[1,2]\**

[1]School of Mechanical Engineering, Tel-Aviv University, Tel-Aviv, 69978, Israel

[2]Department of Biomedical Engineering, Tel-Aviv University, Tel-Aviv, 69978, Israel

*E-mail: gyossifon@tauex.tau.ac.il

**Contents**

**S1. Effective parameters**

The nucleus comprises cytoplasm and a membrane, with the membrane having a finite thickness (~10 nm) and dielectric properties that may differ significantly from those of the cytoplasm. To account for the heterogeneous electrical characteristics of the nucleus, the membrane electrical conductivity ($\sigma_{mem}$) and the conductivity of the fluid inside the membrane ($\sigma_{cyt}$) are replaced by an effective homogeneous conductivity, as given by [1-2]:

$$\begin{cases} \sigma_n = \sigma_{mem}[2(1-\beta)\sigma_{mem} + (1+2\beta)\sigma_{cyt}]/[(2+\beta)\sigma_{mem} + (1-\beta)\sigma_{cyt}] \\ \beta = (1-h/a)^3 \end{cases} \quad (1)$$

where *a* is the radius of the nucleus and *h* is the thickness of the membrane. Notably, this equation is valid only when *h/a*<<1. The effective homogeneous permittivity can be obtained in a similar way:

$$\begin{cases} \varepsilon_n = \varepsilon_{mem}[2(1-\beta)\varepsilon_{mem} + (1+2\beta)\varepsilon_{cyt}]/[(2+\beta)\varepsilon_{mem} + (1-\beta)\varepsilon_{cyt}] \\ \beta = (1-h/a)^3 \end{cases} \quad (2)$$

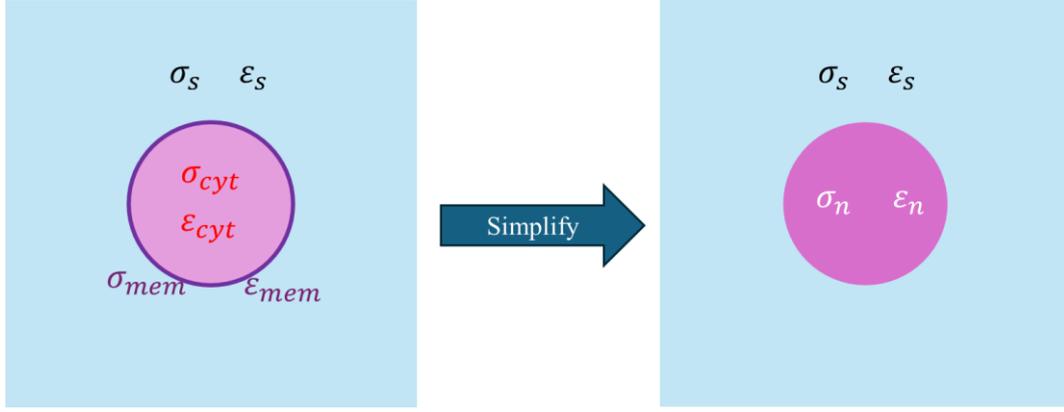

**Figure S1.** Schematic of the effective parameters for the "homogenous" nucleus model.

## S2. Physical properties of the nucleus

In the experiments [3], two types of cell nucleus were used: MDA-MB-231 and MCF-7. Since the dielectric properties and exact dimensions of these specific nuclei are unavailable, we assumed them to be similar and referred to the literature for typical values to be used for both cell types, with surface tension as the only fitting parameter. A range of nuclei dielectric and geometric properties reported previously in the literature are summarized in Table S1. It has been experimentally demonstrated that positive DEP (pDEP) force occurs during the JP-nucleus interaction for both nucleus types at the given frequency of 1MHz [3]. Using this observation as a criterion, we used the values of permittivity and conductivity for the cytoplasm and membrane from various references and calculated the homogeneous effective parameters using Equations 1-2. Furthermore, we calculated the real part of Clausius-Mossotti factor to analyze the DEP response at various frequencies by using following equations:

$$K(\omega) = \left(\frac{\sigma_n - \sigma_s}{\sigma_n + 2\sigma_s}\right)\left(\frac{i\omega\tau_0 + 1}{i\omega\tau_{MW} + 1}\right) \quad (3)$$

$$\tau_{MW} = \left(\frac{\varepsilon_n + 2\varepsilon_s}{\sigma_n + 2\sigma_s}\right) \quad (4)$$

$$\tau_0 = \left(\frac{\varepsilon_n - \varepsilon_s}{\sigma_n - \sigma_s}\right) \quad (5)$$

As shown in Figure S2, some parameter sets ensure that a pDEP force occurs at 1 MHz, while others do not. Since our focus in this study is specifically on the 1 MHz case, we ultimately selected the dielectric parameters from Ref. [4] as representative values for our simulations. But the diameter of the nucleus is set as 10μm based on the measurement in Ref. [3].

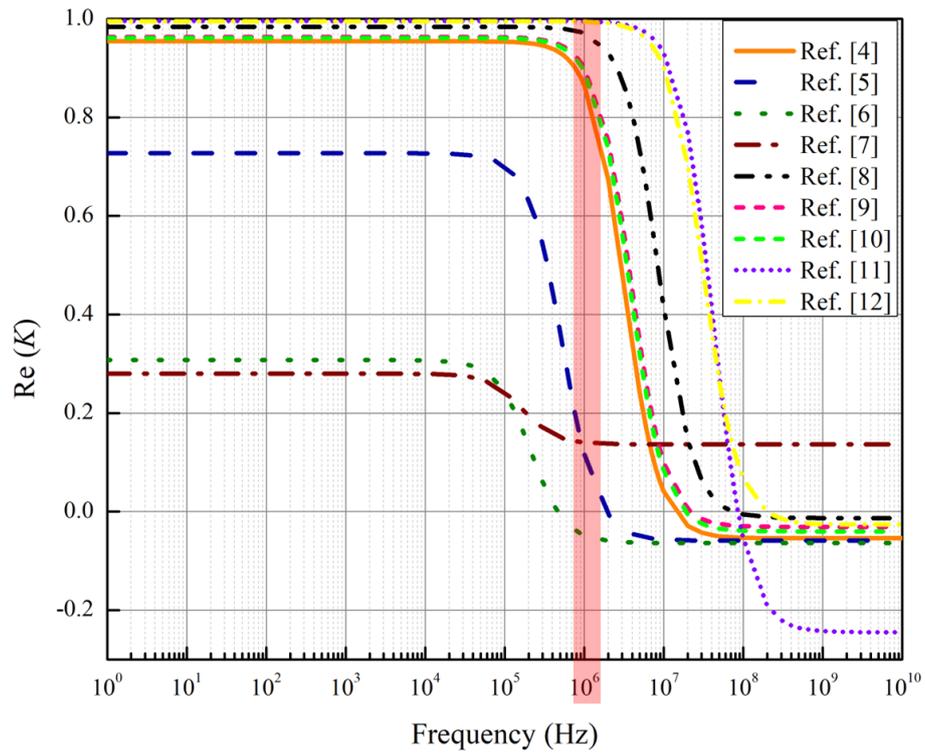

**Figure S2.** The real part of Clausius-Mossotti factor by applying parameters from various references [4-12].

Table S1. Summary of nucleus properties in previous published papers

| Parameters [unit] | Values | | | | | | | | | |
|---|---|---|---|---|---|---|---|---|---|---|
| Nucleoplasm relative permittivity $\varepsilon_{cyt,r}$ | - | 67 | 67 | 67 | 120 | 80 | 72.3 | 70 | 32 | 74 |
| Nucleoplasm conductivity $\sigma_{cyt}$ [S/m] | - | 0.55 | 0.55 | 0.32 | 0.18 | 1 | 1.07 | 0.5 | 2.2 | 1.74 |
| Nuclear membrane relative permittivity $\varepsilon_{mem,r}$ | - | 11.7 | 5 | 11.7 | 22.8 | 10 | 10 | 7 | 32 | 28.4 |
| Nuclear membrane conductivity $\sigma_{mem}$ [S/m] | - | $8.3\times10^{-5}$ | $1.1\times10^{-5}$ | $1.1\times10^{-5}$ | $4.3\times10^{-3}$ | $1\times10^{-3}$ | $1\times10^{-4}$ | $1.0\times10^{-4}$ | $4.5\times10^{-3}$ | $4.2\times10^{-3}$ |
| Nuclear diameter $d$ [μm] | 10 | 2.5 | 5 | 11.8 | 8.8 | 10 | 6 | 7.3 | 2.2 | 8 |
| Nuclear membrane thickness $h$ [μm] | - | 0.01 | 0.01 | 0.04 | 0.04 | 0.04 | 0.01 | 0.01 | 0.04 | 0.04 |
| Reference | [3] | [4] | [5] | [6] | [7] | [8] | [9] | [10] | [11] | [12] |
| Cell type | MDA-MB-231 MCF-7 | General parameters suitable for modeling standard nuclei. | A general spherical biological cell | Lymphocytes | Human Jurkat T lymphocyte cell | A general biological cell | General mammalian cells | B16-F1 Cell | Lymphocytes | Human lymphocyte |

Note: In experiments [3], the electric conductivity and relative permittivity of the solution are $0.6\times10^{-3}$ S/m and 78, respectively.

Finally, Table S2 summarizes the physical properties of the surrounding fluid and the equivalent homogeneous droplet/nucleus, with dielectric properties taken from Ref. [4]. The baseline surface tension of the droplet is set at 0.0002 N/m, as reported in Ref. [13], and serves as the reference for comparisons with other cases.

**Table S2** Physical properties of the equivalent droplet and surrounding fluid

| Fluid | Density (kg/m$^3$) | Viscosity (Pa·s) | Relative permittivity | Electric conductivity (S/m) |
|---|---|---|---|---|
| Nuclues | 1040 [14] | 2×10$^{-3}$ [15] | 66 | 38.5×10$^{-3}$ |
| Solution | 1000 | 1.0×10$^{-3}$ | 78 | 0.6×10$^{-3}$ |

**S3. Boundary Conditions**

In the present study, we focus exclusively on the cases at 1MHz, which is significantly higher than the RC frequency (1.3kHz) of the induced electrical double layer (EDL). Therefore, the metallic side of the JP can be simplified using a floating potential boundary condition, excluding the EDL. However, in general cases where the induced EDL is not negligible, particularly at low frequencies, the metallic side of the JP is often modeled as a capacitor [16]:

$$\mathbf{n} \cdot (\sigma \nabla \tilde{V}) = i\omega C_{DL}(\tilde{V} - \tilde{V}_f) \qquad (6)$$

with capacitance per unit area of $C_{DL} = \varepsilon/\lambda$, where $\varepsilon$ is the permittivity, $\sigma$ is the conductivity, $\lambda$ is the Debye length, $\tilde{V}_f$ is the unknown phasor floating potential determined by imposing the condition of zero total electric flux across the metallic side of the JP, and **n** is the unit normal vector. On the dielectric side of the JP, an insulation boundary condition is imposed:

$$\mathbf{n} \cdot (\sigma \nabla \tilde{V}) = 0. \qquad (7)$$

Figure S3 illustrates the initial electric field distribution surrounding around the JP in the vicinity of a nucleus under an applied AC electric field at varying frequencies, with the induced EDL modeled on the JP's metallic side. At low frequency (10$^3$Hz), the field distribution is relatively uniform due to EDL screening, causing the JP to behave like an insulated particle. As the frequency increases, the EDL screening effect diminishes, and at sufficiently high frequencies (>10$^5$Hz), the metallic side approaches an equipotential state, leading to electric field intensification at the metallo-dielectric interface of the JP.

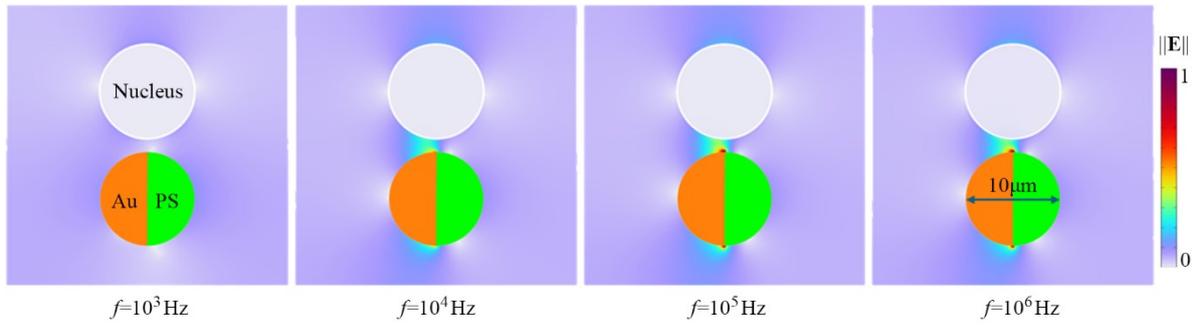

**Figure S3.** The electric field at different frequencies when the induced EDL is considered on the metallic side of the JP.

Figure S4 compares the electric field distribution around a JP interacting with a nucleus at a high frequency of interest ($10^6$ Hz), both with (i.e., using Equation 6) and without (i.e., assuming the metallic side is equipotential) consideration of the electrical double layer (EDL). The top row (t = 0) shows the initial field distribution, while the bottom row (t = 1) illustrates the field after interaction has evolved.

At this frequency of $10^6$ Hz, the results with and without EDL are very similar (the relative error in terms of the dielectric coverage ratio is ~4.5%), indicating that the influence of the EDL can be neglected. The electric field remains highly non-uniform around the JP, with the strongest localization at the JP's interface, driving nucleus deformation and movement. This suggests that at $10^6$ Hz, the polarization response is primarily governed by the bulk dielectric properties of the materials rather than interfacial charge effects, confirming that the floating boundary condition is sufficient to accurately model JP-nucleus interactions at such frequencies.

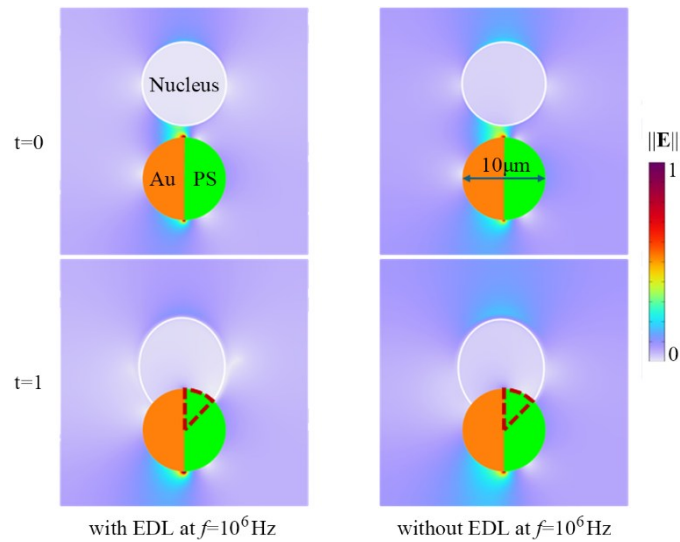

**Figure S4.** The electric field of JP-nucleus interaction at the high frequency of interest (1MHz) with and without induced EDL at t=0 and t=1.

Figure S5 presents the Clausius–Mossotti factor and JP–nucleus interaction, based on parameters from Table S2 and simulations using the floating boundary condition under a 12 V applied voltage. As the frequency increases beyond 1 MHz—where pDEP is observed experimentally—a transition to nDEP occurs, causing the nucleus to migrate toward regions of low electric field strength. The strong agreement between analytical and numerical results confirms that the cycle-averaged force method effectively captures frequency-dependent interaction dynamics, demonstrating the robustness of the proposed modeling approach.

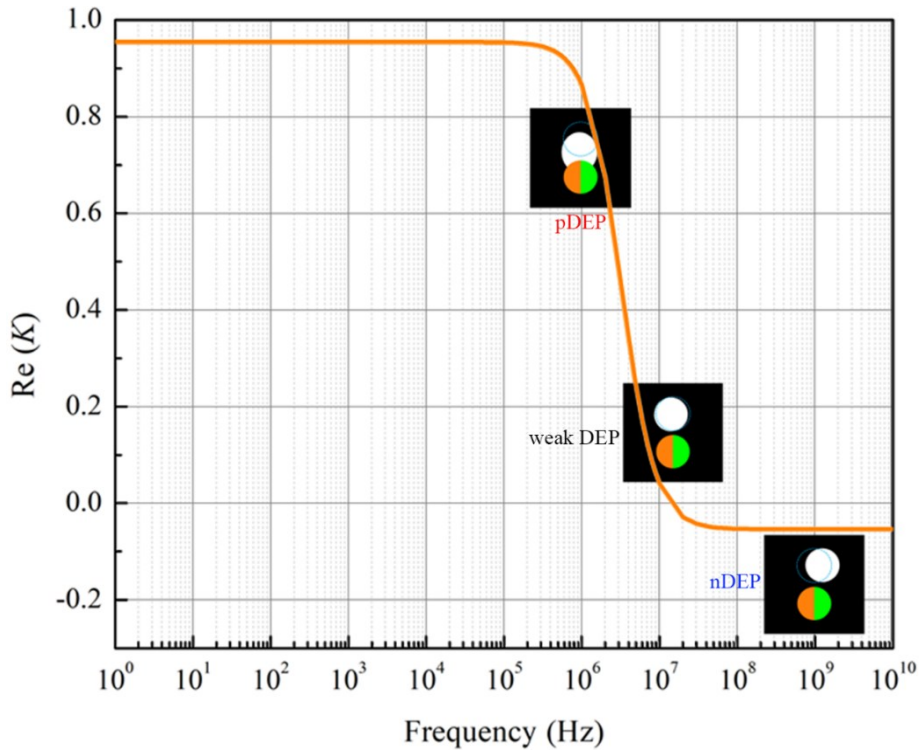

**Figure S5.** The analytical plot of Clausius-Mossotti factor and numerical results of JP-nucleus interactions at different frequencies when $V_{pp}$=12V (the blue line depicts the initial location of the nucleus).

**S4. Grid-independence**

To ensure the grid independence of the simulation results, we conducted a grid convergence study with grids of varying resolutions, as detailed below. While the coarse Grid_1 successfully captures the key features of attraction and deformation of the nucleus on the JP surface, the resolution is relatively low. As the grid resolution increases, Grid_3 and Grid_4 show minimal differences, indicating that the solution has become grid-independent. To balance computational cost and accuracy, we selected Grid_3 for all subsequent simulations and parametric analyses presented in this manuscript.

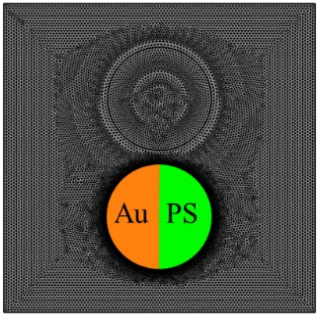

| Grid | Number of cells |
|---|---|
| Grid_1 | 10146 |
| Grid_2 | 19926 |
| Grid_3 | 29678 |
| Grid_4 | 40060 |

Table 3 Grids with different resolutions

**Figure S6.** The computational grid.

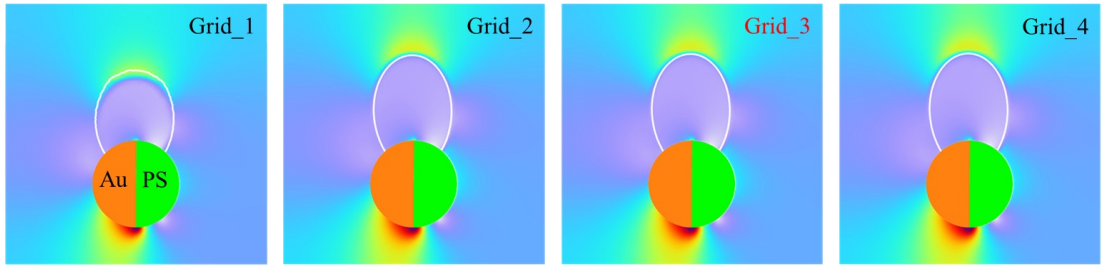

**Figure S7.** Simulation results using different grids under $V_{pp}$=18V, $f$=1MHz.

## S5. Supplemental movie

Movie S1: Depicts the dynamic evolution of the JP–nucleus interaction at a frequency of 1 MHz under an applied voltage of 12V. It clearly demonstrates how the positive dielectrophoretic (pDEP) force induces the attraction and deformation of the nucleus over the JP.